\documentclass[twocolumn]{aastex63}

\def\lta{\mathrel{\spose{\lower 3pt\hbox{$\mathchar"218$}}
     \raise 2.0pt\hbox{$\mathchar"13C$}}}
\def\gta{\mathrel{\spose{\lower 3pt\hbox{$\mathchar"218$}}
     \raise 2.0pt\hbox{$\mathchar"13E$}}}

\newcommand\T{\rule{0pt}{3.2ex}}
\newcommand\B{\rule[-1.2ex]{0pt}{0pt}}

\submitjournal{ApJ}

\shorttitle{2020 broadband activity of OJ 287}
\shortauthors{Kushwaha et al. 2020}

\begin{document}

\title{Blazar OJ 287 After First VHE Activity: Tracking the Re-emergence of the HBL
like Component in 2020}

\correspondingauthor{Pankaj Kushwaha}
\email{pankaj.kushwaha@aries.res.in}

\author[0000-0001-6890-2236]{Pankaj Kushwaha}
\affiliation{Aryabhatta Research Institute of Observational Sciences (ARIES),
Manora Peak, Nainital 263002, India}
\altaffiliation{Aryabhatta Postdoctoral Fellow}

\author[0000-0001-6890-2236]{Main Pal}
\affiliation{Centre for Theoretical Physics, Jamia Millia Islamia, New Delhi 110025, India}

\author[0000-0002-9323-4150]{Nibedita Kalita}
\affiliation{Key Laboratory for Research in Galaxies and Cosmology, Shanghai Astronomical Observatory, Chinese Academy of Sciences, 80 Nandan Road, Shanghai 200030, China}

\author[0000-0003-0071-8947]{Neeraj Kumari}
\affiliation{Astronomy and Astrophysics Division, Physical Research Laboratory, 
Navrangpura, Ahmedabad 380009, Gujarat, India}
\affiliation{Indian Institute of Technology, Gandhinagar 382355, Gujarat, India}

\author[0000-0003-2865-4666]{Sachindra Naik}
\affiliation{Astronomy and Astrophysics Division, Physical Research Laboratory,
Navrangpura, Ahmedabad 380009, Gujarat, India}

\author[0000-0002-9331-4388]{Alok C Gupta}
\affiliation{Key Laboratory for Research in Galaxies and Cosmology, Shanghai Astronomical
Observatory, Chinese Academy of Sciences, 80 Nandan Road, Shanghai 200030, China}
\affiliation{Aryabhatta Research Institute of Observational Sciences (ARIES),
Manora Peak, Nainital 263002, India}

\author[0000-0001-8058-4752]{E.~M. de Gouveia Dal Pino}
\affiliation{Department of Astronomy (IAG-USP), University of Sao Paulo, Sao Paulo
05508-090, Brazil}

\author[0000-0002-4455-6946]{Minfeng Gu}
\affiliation{Key Laboratory for Research in Galaxies and Cosmology, Shanghai Astronomical Observatory, Chinese Academy of Sciences, 80 Nandan Road, Shanghai 200030, China}

\begin{abstract}
We report the re-emergence of a new broadband emission through a detailed and systematic
study of the multi-wavelength spectral and temporal behavior of OJ 287 after its
first-ever reported VHE activity in 2017 to date, which includes the 
second-highest X-ray flux of the source. The source shows high optical to X-ray
flux variations, accompanied mainly by strong spectral changes. The optical to X-ray 
flux variations are correlated and simultaneous except for two durations when they
are anti-correlated. The flux variations, however, are anti-correlated with the X-ray spectral state while correlated with optical-UV (ultraviolet). Weekly binned {\it Fermi}-LAT data around the duration of the highest
X-ray activity show a few detections with a log-parabola model but none with a
power-law; yet the extracted LAT spectral energy distribution (SED) of the high activity duration for both the models is similar and show a hardening above 1 GeV. Further, near-infrared (NIR) data indicate strong spectral change, resembling a thermal component. Overall, the combined optical
to gamma-ray broadband spectrum establishes
the observed variations to a new high-energy-peaked (HBL) broadband emission component, similar to the one seen during the highest reported X-ray flux state
of the source in 2017. The observed activities indicate some peculiar
features that seem to be characteristic of this emission component while its
appearance, a few years around the claimed
\(\sim 12\)-year optical outbursts strongly indicate a connection between the two.
\end{abstract}


\section{Introduction} \label{sec:intro}
Blazars are the most prominent and persistent non-thermal emitters with a continuum
powered by a relativistic jet roughly aligned with our line of sight. The continuum
shows variations on all accessible timescales and spread across the entire accessible
electromagnetic (EM) spectrum. Detailed multi-wavelength (MW) studies, especially since
the launch of the Fermi observatory, have revealed that the variability is pervasive 
-- seen in all: flux, spectra, and polarization with time with changes in one often
accompanied by changes in others. These studies
have also firmly established or confirmed many phenomenological features e.g. the
variations are stochastic \citep[e.g.][]{2014ApJ...786..143S}, the broadband emission show a characteristic bi-modal
spectral energy distribution (SED) with one peak in-between infra-red to UV/X-ray energies and
the other at MeV/GeV energies \citep[e.g.][]{2010ApJ...716...30A}. Interestingly, despite being extreme among the accretion-powered sources and
a jet-originated continuum, in terms of temporal flux variability, their long-term
variations broadly exhibits the statistical behavior shown by accretion-powered 
sources \citep[e.g. ][and references therein]{2015SciA....1E0686S,2016ApJ...822L..13K,2017ApJ...849..138K}.

The most unique and characteristic observational feature of blazars' highly variable
broadband emission is the broad bi-modal SED extending
from the lowest accessible EM band i.e. the radio to the highest accessible i.e.
GeV-TeV \(\gamma\)-rays. The broadband SED of all blazars can be clubbed
into three different spectral sub-classes: low-energy-peaked (LBL/LSP), 
intermediate-energy-peaked (IBL/ISP), and high-energy-peaked \citep[HBL/HSP;][]
{1998MNRAS.299..433F,2010ApJ...716...30A} based on the location of the low-energy
hump. A remarkable property of each spectral sub-class is the stability
of the location of the two peaks despite huge variations in flux and often 
spectral shape.  Only in a few rare instances, an appreciable shift in the location
of the peaks has been observed e.g. the 1997 outburst of Mrk 501 
\citep{1998ApJ...492L..17P,2018A&A...620A.181A} and the end-2015 to the mid-2017
activity of OJ 287 \citep{2018MNRAS.473.1145K,2018MNRAS.479.1672K}. Even these two
cases are remarkably different. In the case of Mrk 501, the location of both the
peak shifted to higher energies. On the contrary, in OJ 287, a shift in the location
of only the high-energy peak was observed during the 2015 -- 2016 activity 
\citep{2018MNRAS.473.1145K, 2018bhcb.confE..22K} while in the 2016 -- 2017, a new
broadband emission component overwhelmed the overall emission, appearing
as an overall shift in both the peaks as revealed in the detailed study by
\citet{2018MNRAS.479.1672K}.
The SED being the prime observable for exploration of the yet debated high-energy
emission mechanisms, such changes offer invaluable insights about the emission processes.
For example, in the Mrk 501, the shift in both peaks strongly imply the same
particle distribution for the overall emission while for OJ 287, the shift of only
the high-energy peak can be reproduced either by inverse Compton scattering
of the broad-line region photon field \citep{2018MNRAS.473.1145K}
or emission of hadronic origin \citep{2019MNRAS.489.4347O,2020MNRAS.498.5424R}.

In terms of spectral variations and spectral state, OJ 287 is the most dynamic 
blazar known so far \citep[and references therein]{2001PASJ...53...79I,2017MNRAS.468..426S,
2017IAUS..324..168K,2018MNRAS.479.1672K, 2018MNRAS.473.1145K,2020ApJ...890...47P,
2020Galax...8...15K,10.1093mnrasl}. Contrary to a few of the blazars that have
shown a very different SED compared to the SED considered normal of them and
is the basis of their spectral classification in the blazar spectral sequence once or twice \citep[e.g. Mrk 501;][]{1998ApJ...492L..17P,
2018A&A...620A.181A},
OJ 287 is been in a different broadband SED states since its 2015
activity to date \citep{2018MNRAS.473.1145K,2018MNRAS.479.1672K,2013MNRAS.433.2380K}. It belongs to the BL Lacertae subclass
of blazars -- the blazar subclass characterized by a lack of strong or completely
absent emission line features in their optical spectrum. Located at a redshift
of $z=0.306$ \citep{1985PASP...97.1158S,2010A&A...516A..60N}, it is one of the
most bright and dynamic sources at optical and radio energies. In addition to exhibiting
typical stochastic variability of blazars \citep[e.g.][]{2018ApJ...863..175G}, it is
the blazar with the highest number of claims of QPO \citep[e.g.][]{1973ApJ...179..721V,
1985Natur.314..148V,1988ApJ...325..628S,2013MNRAS.434.3122P,2016ApJ...832...47B,2018ApJ...866...11D,2020MNRAS.499..653K} with some persistent
ones, 
notably a recurring \(\sim 12\)-yr  optical outbursts \citep[QPOs;][and references therein]{1988ApJ...325..628S,2018ApJ...866...11D},
a probable \(\sim 22\)-yr periodicity structure in the motion of bright radio
knots and an yearly modulation in the location of quasi-stationary knots
\citep{2018MNRAS.478.3199B} as well as a \(\sim 30\)-yr periodic feature
\citep{2017Galax...5...12C}, that have made it an excellent source to explore various 
aspects of active galactic nuclei physics \citep[e.g.][and references therein]{2011ApJ...726L..13A,2018ApJ...864...67S,2018ApJ...866...11D,2018MNRAS.478.3199B,2020Galax...8...15K}.

As per broadband SED, OJ 287 is a low-frequency-peaked/low-synchrotron-peaked (LBL/LSP)
blazar with a low-energy peak around near-infrared (NIR) energies and the high-energy
peak around 100 MeVs \citep{2010ApJ...716...30A,2013MNRAS.433.2380K}. However, since
the start of a dynamic MW activity in mid-November 2015, studies have reported 
drastic spectral variations in optical to MeV-GeV bands and very different and 
distinct broadband spectral phases \citep{2017IAUS..324..168K,
2018MNRAS.473.1145K,2018MNRAS.479.1672K,2018MNRAS.480..407K, 2020Galax...8...15K,
10.1093mnrasl} compared to its well-known LBL SED. The associated
temporal flux variations were also different with lags and leads compared to generally
seen simultaneous MW variations. The overall temporal and spectral activity
during this duration can be categorized into two episodes. While the first activity
episode from mid-November 2015 to May 2016 was temporally similar to its typical
reported behavior -- showing simultaneous MW variability, the broadband spectrum
was very different; with two new changes -- a break in the NIR-optical spectrum
from its well-known (smooth) power-law form and a shift in the location of the high-energy
peak to higher energies \citep{2018MNRAS.473.1145K}. The NIR-optical break was found
to be consistent with a standard accretion-disk emission of a \(\sim 10^{10}~M_\odot\)
black hole while the shift in high-energy peak can be explained by inverse Compton scattering of broad line photons \citep{2018MNRAS.473.1145K}.
The explanation of NIR-optical break being a standard accretion-disk emission,
if true, has direct implications to another ongoing debate about the central
engine of the source invoked to explain the recurrent \(\sim 12\)-yr optical outbursts.
Both, the mass of the central SMBH inferred from the NIR-optical explanation and the
temporal coincidence
of its appearance currently favor the binary disk-impact model \citep[and references
therein]{2020Galax...8...15K}.

The second MW activity immediately followed the first phase in mid-2016 and continued
till mid-2017 \citep{2017IAUS..324..168K,2018MNRAS.479.1672K,10.1093mnrasl}. Contrary
to the activity in the first phase, it was unique spectrally and temporally. While
the source reached the strongest yet reported optical to X-ray emission \citep{2017ATel10043....1G}, it was rather steady with a low level of emission despite
detection on daily timescale in the LAT energy band \citep{2018MNRAS.479.1672K}. A 
detailed study by \citet{2018MNRAS.479.1672K} revealed that at the beginning of the
activity, optical was leading X-ray while the variations were simultaneous in-between
and finally ended with X-ray leading the optical. Tracking of spectro-temporal 
evolution further revealed that the observed variability was driven
by a new high-energy-peaked emission component, corroborated also by the first-ever
detection of the source at very high energies \citep[VHEs, E\(> 100\) GeVs;][]
{2017ICRC...35..650B}. 

 In 2020 also, the source again exhibited a very high optical to X-ray activity
and achieved the second brightest reported X-ray flux \citep{2020ATel13658....1K,
10.1093mnrasl}. The optical to X-ray flux evolution show strong changes driven by
a very soft X-ray spectral state \citep{10.1093mnrasl}, similar to the one seen
during the 2017 high optical to X-ray activity. The continued follow-up of the 2020
activity revelaed, for the first time, a signature of iron absorption in the
XMM-Newton data, though it was absent in the NuSTAR data taken 10 days after the
XMM-Newton observation. 

Here, we present a detailed exploration of spectro-temporal evolution and
spectral changes from optical to LAT gamma-ray energies between May 27, 2017 to
June 10, 2020 (MJD: 57900 -- 59010) and compare the behaviors with the 2017
activity that was also driven by the soft X-ray spectrum. The period covered
here corresponds to the end-phase of the first VHE activity and the highest recorded
X-ray flux state to the end of the second highest reported X-ray flux. This
allows a continuous access to the flux and spectral behavior of the source and
in turn, to the evolutionary track of the new broadband emission component that
was responsible for the 2017 activity and its revival, resulting in the 2020
activity. The work is presented
in five sections with the next section describing the data sources and reduction
procedures. In section \S3, we explore the spectral and temporal behavior with a
comparative discussion in \S4. Finally we summarize and conclude in \S5.

\section{MW Data Reduction} \label{sec:data}
 The MW data used in this work are taken from the public archives
of different observatories. The optical to X-ray data is mainly taken from the
{\it Swift} observatory and a study focusing on optical to X-ray flux and
X-ray spectral variations is already
presented in \citet{10.1093mnrasl} \citep[a part of long-term monitoring program of OJ 287;][]{2017IAUS..324..168K}. Our focus is on optical to X-ray spectro-temporal
evolution and broadband spectral changes and thus, we reanalyzed the {\it Swift} data and supplemented these with the \(\gamma\)-ray data from the {\it Fermi}-LAT (Large Area Telescope) during the high optical to X-ray flux state. In addition, we also
attempted observations at optical from ARIES and NIR from Mount Abu Infrared
Observatory (MIRO) at Mt Abu, Rajasthan, India \citep{2012BASI...40..243B} after the
report of increased activity at optical energies \citep{2020ATel13755....1H} following
the X-ray peak. However, due to adverse observing conditions, we could only observe
the source at NIR from MIRO. 

\textbf{Swift-UVOT:} We used the level-2 sky images of all the \textit{Swift}-UVOT
\citep[Ultra-Violet Optical Telescope; ][]{2005SSRv..120...95R} filters (V,
B, U, UVW1, UVM2, UVW2) to extract the
magnitude/flux density. For this, we used the {\tt uvotsource} task available
within HEASOFT package (version 6.28) with a source region
of 5$^{''}$ and an annular region for the background (free of any source) with an inner
and outer radius of 10$^{''}$ and 20$^{''}$, respectively. Observations
in which the source location was on bad pixels were discarded. We used the updated
callibration file that also account for the loss of sensitivity of the UV detector
with time\footnote{\url{https://www.swift.ac.uk/analysis/uvot/index.php}}. Finally,
the extracted flux densities were corrected for interstellar reddening using an 
E(B-V) = 0.0241 \citep{2011ApJ...737..103S}.

\textbf{Swift-XRT:} We only used the pointing mode data taken
in the photon counting (PC) and window timing (WT) mode by the \textit{Niel Gherel 
Swift Observatory} \citep{2005SSRv..120..165B} and followed the standard
recommended data reduction procedures. Firstly, each observation was
reprocessed using the standard default filtering criterion and latest calibration
files (updated on 20200504) with the \textit{xrtpipeline} (version 0.13.5) task. Then for each observation, we generated a source spectrum from a circular 
region of 47$^{''}$ and a background spectrum from an annular region with inner
and outer radius of 80$^{''}$ and 176$^{''}$ using the \textit{xselect} task.
We also checked for pile-up in the PC\footnote{None of the WT mode data has source
region count rate $\gtrsim 100$ counts/s when pile-up is expected} mode data whenever the source count rate exceeded
0.5 counts/s and corrected following the recommended procedure\footnote{\url{http://www.swift.ac.uk/analysis/xrt/pileup.php}} by discarding
the central 2-4 pixels. The corresponding ancillary response files
were generated through the {\tt xrtmkarf} task. Finally, we binned the source
spectrum files with a minimum count of one per bin using the {\tt grppha} tool
for spectral fitting.

We performed the spectral fitting in {\tt XSPEC} (version 12.11.1) employing
\textit{Cash statistic} following the approach adopted in \citet{2018MNRAS.479.1672K}.
We used both power-law (PL: \(N_p \propto E^{-\Gamma}\)) and log-parabola model (LP:
\(N_p \propto E^{-\alpha-\beta log(E)}\)) and the best-fit model was chosen based
on the \textit{F-test} probability. The LP model was selected as the best-fit
model if the \textit{F-test} probability was $\rm \leq 0.05$. In the fitting process,
all the parameters were free to vary initially. The hydrogen column density ($n_H$) value was fixed to the Galactic
value of \(2.38 \times 10^{20}\) cm$^{-2}$ \citep{2016A&A...594A.116H} whenever the fit value was below
the Galactic value or consistent with it. The best-fit model was used to estimate
the unabsorbed flux in 0.3 -- 10.0 keV band using the {\tt cflux} task.

We verified our result using \textit{chi-squared} statistics by binning the 
spectra with 20 counts per bin and found that both results are consistent
except when the counts are low allowing only fewer bins\footnote{ While fitting
with all parameters free to vary initially, in
some cases, despite good exposure and photon counts (similar to other Ids),
we got large errors in the derived flux values, mainly due to data quality at
low energies. In such cases, we first tried a lower fit energy value of 0.5
keV instead of 0.3 keV. If this failed, we fixed the $n_H$ to the Galactic
value and chose the lower fit range again to 0.3 keV. If all these failed,
we simply fitted a PL model between 0.3 -- 10 keV and derived the fluxes.}.

\textbf{NuSTAR:}
NuSTAR \citep{2013ApJ...770..103H} observed OJ 287 on May 4, 2020 (MJD 58973)
and we reanalyzed the data presented in \citet{10.1093mnrasl}, mainly for joint
fitting with optical to X-ray data from {\it Swift} observatory (ref.
\ref{subsec:joint}). To reduce the observational data, we followed the standard reduction
procedure by using the NuSTAR Data Analysis Software ({\tt NUSTARDAS}). We used the
{\tt nupipeline} (version 0.4.8) task for data reduction with the latest available calibration
files (CALDB no. 20200813). For generating the spectra from both the FPMA and FPMB modules, we selected
a circular region of 50$^{''}$ keeping the source at the center, and also we used
a circular region of 100$^{''}$ for background away from the source. We produced
the spectra and corresponding response files using the {\tt nuproducts} tool available within the
{\tt NUSTARDAS}. We grouped the spectrum with a minimum one count per bin using the {\tt
grppha} tool.

\textbf{Fermi-LAT:}
We used the LAT \citep{2009ApJ...697.1071A} data processed with the latest
PASS-8 instrument response function (P8R3) during the high X-ray activity duration
of the source (MJD: 58785.0-59010.0) and analyzed it following the standard
analysis procedure\footnote{\url{https://fermi.gsfc.nasa.gov/ssc/data/analysis/scitools/python_tutorial.html}} using the data reduction package {\it Fermitools}
(version 1.0.1). For each time bin of interest, we selected the ``SOURCE''
class events ($\rm evclasss=128$) from a circular region of $15^\circ$ (ROI) centered
on the source. Further, we restricted the energy range to within 0.1 -- 300 GeV and
applied the standard zenith angle cut of $90^\circ$. Following these 
selections, we generated good time intervals (GTIs) using the standard expression: 
``(DATA\_QUAL$>$0)$\&\&$(LAT\_CONFIG==1)''. We then generated the exposure map on
the ROI and an additional annular region of $10^\circ$ around it. Finally, to extract the relevant physical quantities like photon flux, spectrum, and energy flux, we performed the
likelihood fitting using the ``unbinned maximum likelihood'' method ``gtlike'' incorporated in the data reduction package with a source model XML file generated
from the fourth Fermi-LAT point source catalog \citep[4FGL;][]{2020ApJS..247...33A}.
The XML model file included the Galactic and extra-galactic contributions through
the respective template files ``gll\_psc\_v21.fits'' and ''iso\_P8R3\_SOURCE\_V2\_v1.txt`` respectively.

Firstly, we performed a fit to 0.1--300 GeV of LAT data for the whole duration\footnote{we did not find any photon of E$>100$ GeV.} using the default
source model LP for the source as in the 4FGL catalog following the iterative approach
adopted in \citet{2014ApJ...796...61K} by removing insignificant sources -- measured
by test Statistics (TS) $\leq$ 0. This best-fit
XML model file was then used for the light curve and SED extraction over the relevant
time bins. Though LP is the default model for the source in 4FGL, our analysis of
monthly binned LAT data presented in \citet{2020MNRAS.499..653K} shows that, in general,
the PL model adequately describes the source spectrum fairly well. On the other hand,
studies of LAT data presented in \citet{2018MNRAS.479.1672K} and \citet{2018MNRAS.473.1145K} show that the source has shown depature from a simple PL spectrum at LAT energies,
especially after its 2015 activity and during 2016--2017 high optical to X-ray activity.
Since the X-ray activity during the current duration is similar
to the 2016--2017 activity of the source, we used both LP and PL for the light curve and SED extraction.
For gamma-ray SED, we divided the 0.1-300 GeV range into six logarithmically equi-spaced
energy bins: 0.1--0.3, 0.3--1, 1--3, 3--10, 10--100, and 100-300 GeV and used
both PL and LP to extract the energy flux. For both light curve and SED, only
data points with TS $\geq$ 9, corresponding to \(\geq~3\sigma\) detection were used
for the scientific analysis.

\begin{figure*}%[h]
\centering
\includegraphics[scale=0.65]{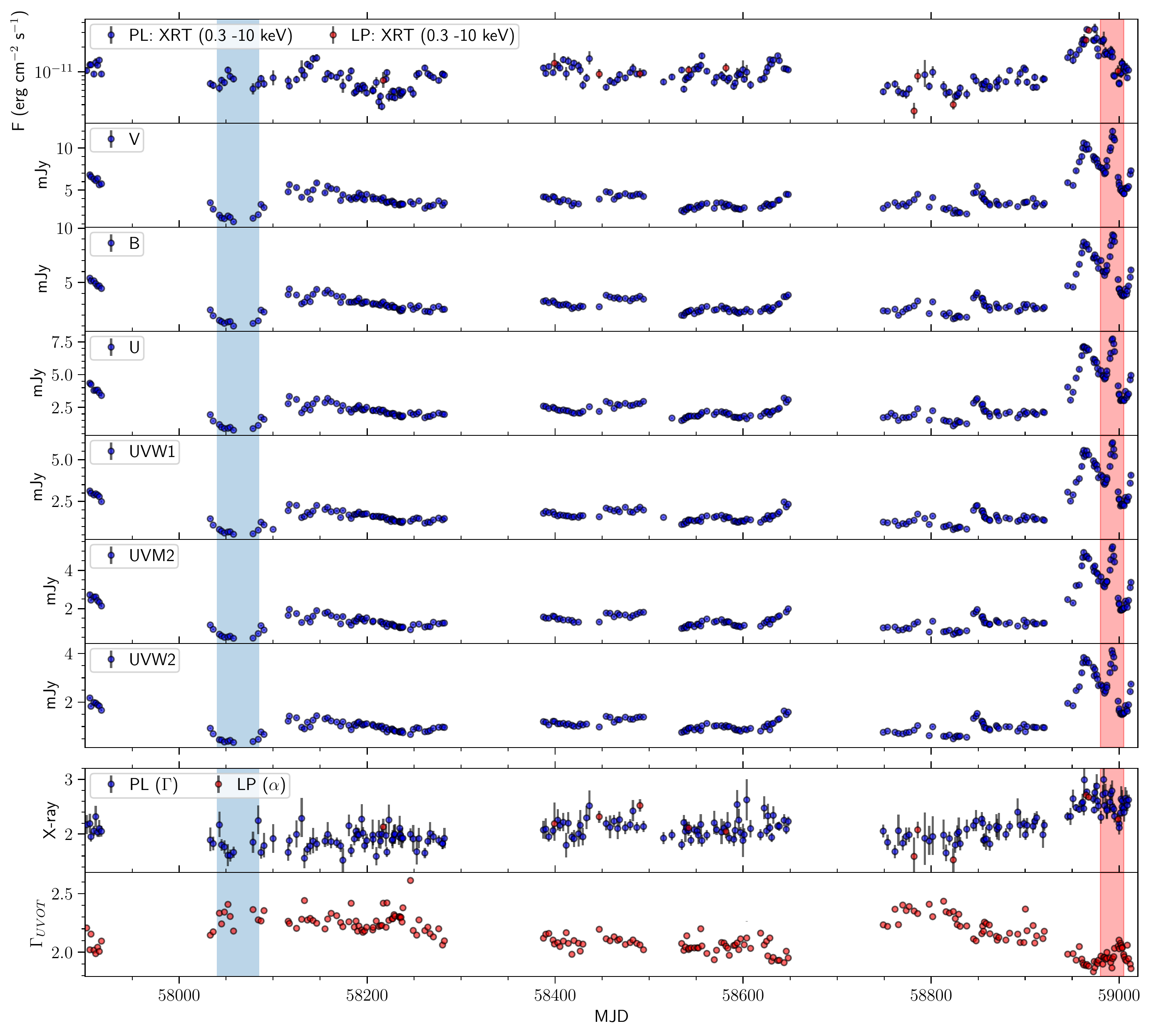}
\caption{{\it Top Panel}: MW light curve of OJ 287 from X-ray to optical energies from
May 27, 2017 to June 10, 2020 (MJD: 57900 -- 59010). Please note the log-scale of 
X-ray flux. {\it Bottom Panel}: Spectral indices for X-ray and UVOT. For X-ray we used both a powerlaw (blue) and a log-parabola
(red) while for UVOT we used powerlaw only. The shadowed regions mark the duration
when the X-ray evolution is at odd with the optical-UV (see \ref{subsec:temporal}).
Similar odd trends were also seen during the 2017 activity \citep[Table 1 and associated
texts; see also \citet{10.1093mnrasl}]{2018MNRAS.479.1672K}.}
\label{fig:mwlc}
\end{figure*}

\textbf{MIRO:}
The NIR observations of OJ 287 were carried out in the $J$ and $K_s$
bands using the Near-infrared Camera and Spectrograph (NICS) instrument on the PRL's
1.2m, f/13 telescope in the imaging mode on 4 June 2020 (MJD 59004). The NICS provides $\sim$1
arcsec seeing over an un-vignetted field of view of 8$\times$8 arcmin$^2$ with 
1024$\times$1024 pixels array configuration. 
% The photometric observations at MIRO were carried out on 4 June 2020 in photometric
% sky conditions in imaging mode.

A set of five frames of the source along
with field stars were taken in $J$ (1.17--1.33 $\mu$m, centered at 1.25$\mu$m) and
$K_s$ (1.99--2.310 $\mu$m, centered at 2.15 $\mu$m) filters at five dithered positions
with exposures of 40~s and 25~s, respectively. The data were reduced by using the {\tt
IRAF} package \citep{1986SPIE..627..733T,1993ASPC...52..173T} as described in \citet{2010MNRAS.404..367N}. In the beginning, the
sky-frame was generated using median-combine of all raw and dithered images and then
subtracted from raw images to get the clean images. Aperture photometry was done
using the {\tt PHOT} task of {\tt IRAF}. Six nearby field-stars were used as the
standard stars to get the apparent magnitude of the target star OJ 287. In this process,
the $J$ and $K_s$-band magnitudes of OJ 287 were estimated to be 12.911$\pm$0.014
and 12.388$\pm$0.018, respectively. The reddening corrected magnitude was then
converted to flux using the zero point flux given in \citet{2003AJ....126.1090C}.

The correction of observed X-ray fluxes for the galactic absorption and the
de-redenning of the optical to UV fluxes leads to increment in the flux values. However,
the mentioned hydrogen column density affects only the spectrum (fluxes) below 1 keV
while the de-reddening of optical-UV fluxes modify the flux at most by a factor of
1.07 to 1.24. As we show in the later part of this work, these changes are negligible
compared to reported flux changes at optical-UV and X-ray energies ($\gtrsim$ 10 between the minimum and the maximum) and thus, has no bearing at all on the HBL
component claim and our findings.

\section{MW Variations}\label{sec:analysis}

\subsection{Temporal}\label{subsec:temporal}
\begin{figure}%[!htba]
  \centering
 \includegraphics[scale=0.50]{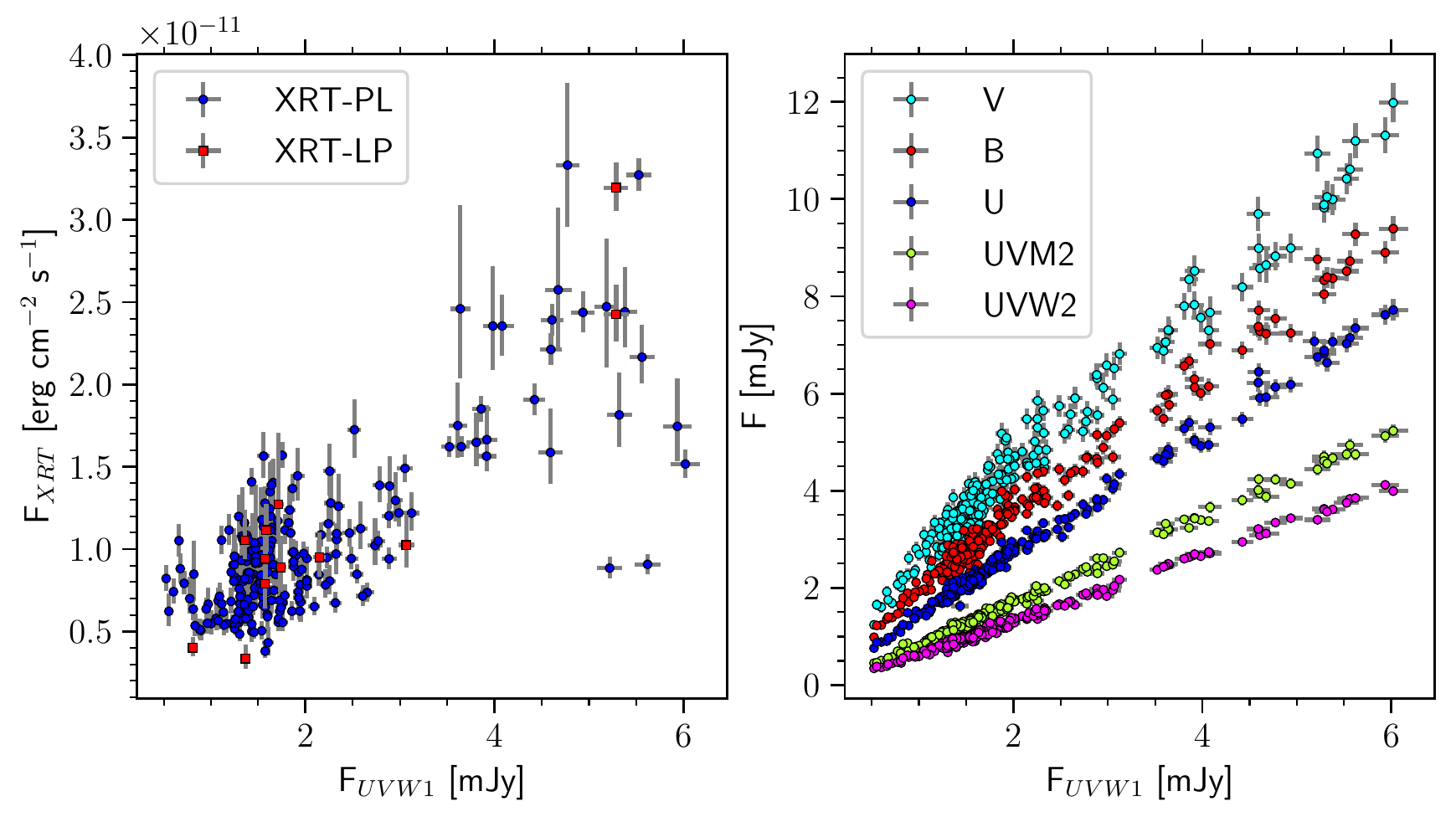}
\caption{Variation of X-ray 0.3--10 keV flux (left panel) and optical-UV (right
panel) with respect to the Swift-UVOT UVW1 band flux. The flux points in the
bottom right corner of the left panel, appearing as outliers, correspond to the
red-shaded duration marked in Figure \ref{fig:mwlc} when UVOT is flaring but XRT
is steady.}
\label{fig:fig2}
\end{figure}

The MW optical to X-ray emission for the duration May 27, 2017 to June 10,
2020 (MJD: 57900 – 59010) is
shown in Figure \ref{fig:mwlc}. Except for the two durations shown by shadowed regions
in Figure \ref{fig:mwlc}, visual inspection shows that the optical to X-ray variations
are simultaneous within the observational cadence. An interesting aspect is that
most of the flux variations are associated with changes in the spectral index
in the optical-UV\footnote{we assumed a simple PL model for optical-UV spectrum} and the
X-ray bands. Also, the spectral changes in optical-UV are
anti-correlated with X-ray -- a hardening at optical-UV is associated with a
softening of X-ray spectrum and vice-versa. This is also apparent from Figure
\ref{fig:fig2} with optical-UV following each other linearly while X-ray, though
shows correlated flux variation, the trend is more complex than linear.

Figure \ref{fig:lat} shows the weekly light curve from \emph{Fermi}-LAT in 0.1--300
GeV band along with the spectral index \(\alpha\) of the LP model and the SEDs
from both the spectral models (PL and LP).
We used both PL and LP model for the light curve and SED extraction to accommodate
spectral changes (ref. \S\ref{sec:data}). However, the two models for OJ 287 resulted
in contrary outputs. While there are a few detections (TS \(\geq\) 9)
with the LP model, the PL model resulted in none. Looking at the SED from
both the model (Fig. \ref{fig:lat} bottom panel), the non-detection with a PL
model could be due to the source being weaker at LAT and the spectral changes.
Further, based
on LP spectral index, a clear hardening of spectral shape is 
apparent for the duration when the X-ray flux started rising (MJD 57875 onward)
leading to the second brightest reported X-ray flux \citep{10.1093mnrasl}.

\subsection{MW SEDs}\label{subsec:SEDs}

\subsubsection{Fermi-LAT Spectrum}\label{subsec:LAT}
\begin{figure}%[ht]
\centering
 \includegraphics[scale=0.6]{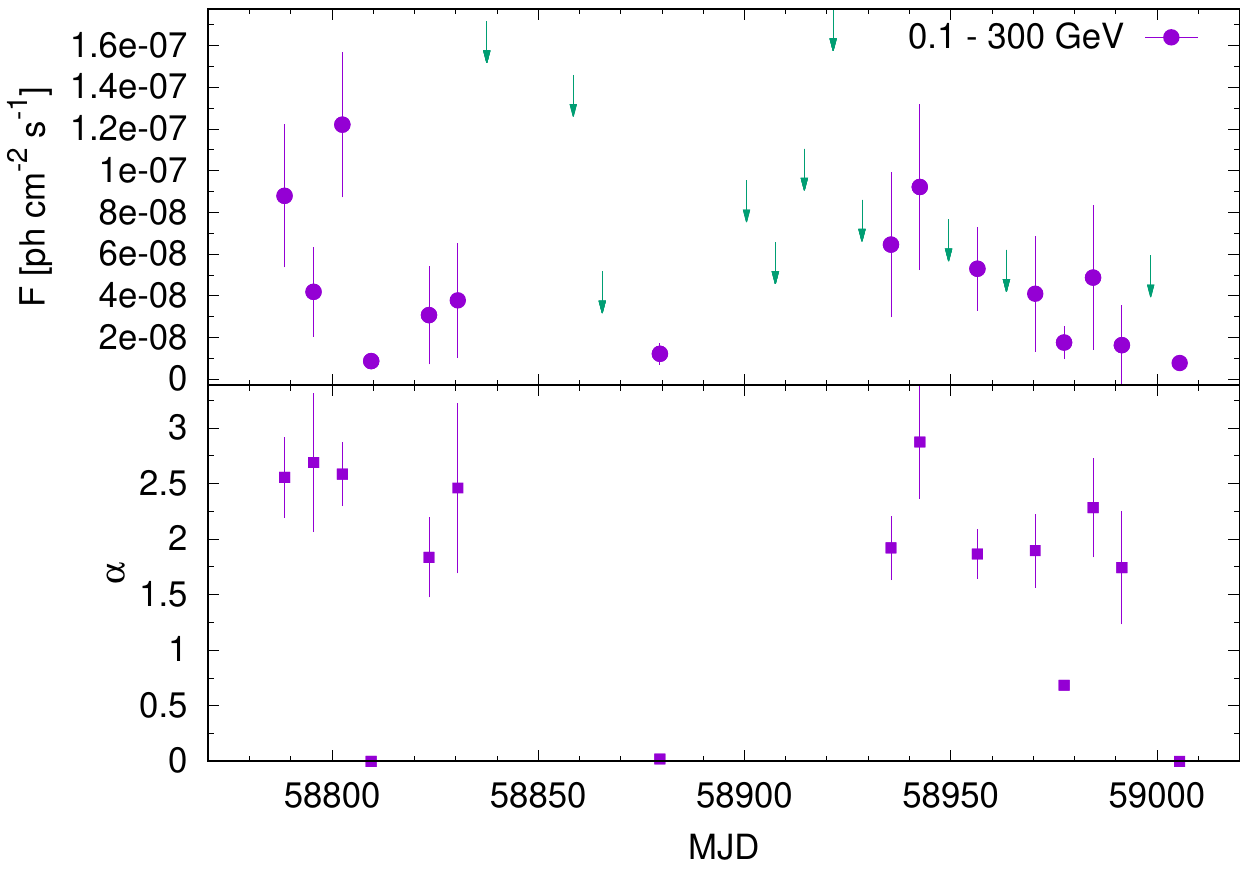}
 \includegraphics[scale=0.63]{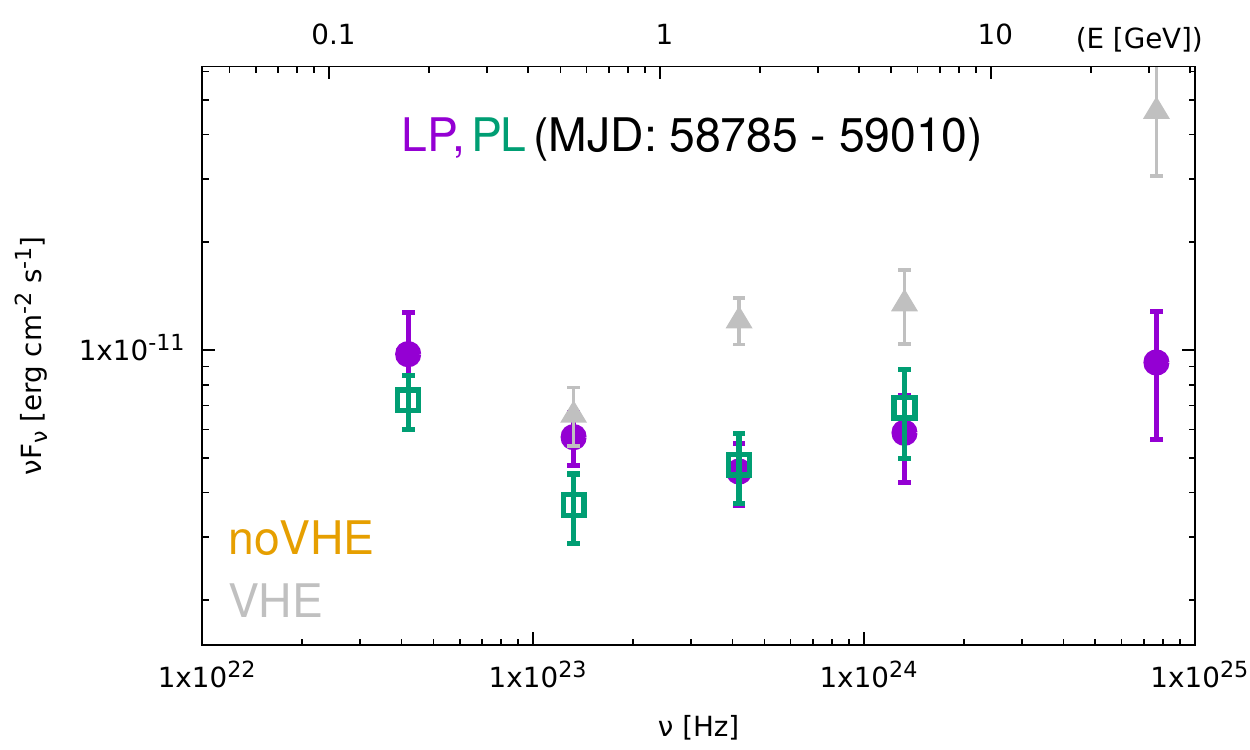}
\caption{{\it Top}: Weekly light curve of OJ 287 focusing around the strong X-ray 
activity duration from MJD 58785 to 59020 along with the LP spectral index $\alpha$.
The downward arrows represent the 95\% (\(2\sigma\)) upper limit.
{\it Bottom:} The MeV-GeV \(\gamma\)-ray spectrum for the duration of the light
curve using a PL (green) and a LP (purple) fit to individual energy bin. For 
comparison, we have plotted the LAT SED from its 2016 -- 2017 activity that was also
due to an extremely soft X-ray spectral state of the source. The orange data labeled
as ``noVHE'' correspond to LAT spectrum before OJ 287 was detected at VHE while
the grey data labeled as ``VHE'' refer to the LAT spectrum from the VHE
activity duration (ref \ref{subsec:LAT} for details).}
\label{fig:lat}
\end{figure}

As mentioned above, the source is too weak even on a week binning of the LAT data.
Thus, we used the whole duration around the high X-ray activity (MJD: 58785 -- 59010)
and extracted the LAT SED in six bands: 0.1--0.3, 0.3--1, 1-3, 3-10, 10-100, and
100-300 GeV. The resulting LAT SEDs for both the spectral models (PL, LP) are
shown in the bottom panel of Figure \ref{fig:lat}. The SED extracted using both
the models are consistent with each other and so is the observed trend in the spectrum
except that with the LP model, OJ 287 is also detected in the 10--100 GeV band
(the highest energy data point). There is a clear spectral hardening above
\(\sim 1\) GeV. Combined with the optical-X-ray
spectrum (ref Fig. \ref{fig:fig4}), it indicates a new broadband emission component,
similar to the one seen during its highest X-ray activity duration in 2016 -- 2017
\citep{2018MNRAS.479.1672K}. For comparison, the LAT spectra from the soft X-ray
spectrum dominated 2016--2017 MW before and during the VHE
detection from \citet{2018MNRAS.479.1672K} is also shown. It clearly shows
that the source is relatively weaker at MeV-GeV energies in the current episode.

\subsubsection{Optical to X-ray Spectral Evolution and MW SEDs}
\begin{figure*}%[ht]
\centering
 \includegraphics[scale=0.75]{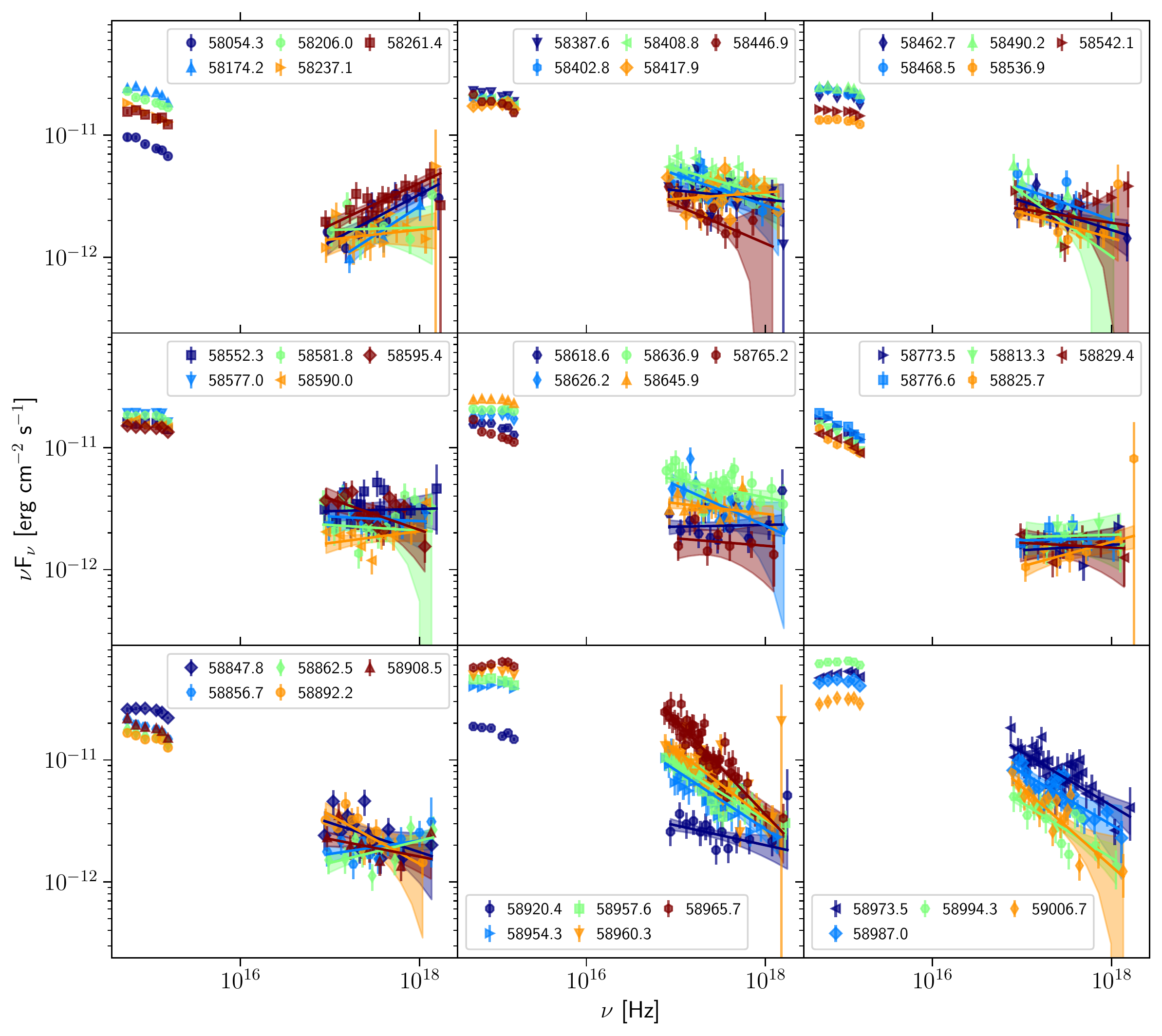}
\caption{Optical to X-ray SEDs of OJ 287, representative of spectral evolution
shown during the considered duration. The solid lines are the best-fit model to the 
X-ray data while the shaded regions are the 1\(\sigma\) range. Only SEDs showing
spectral changes either in optical-UV or X-ray or in both
the bands compared to the preceding SED are shown
 (ref Fig. \ref{fig:mwlc}).}
\label{fig:fig4}
\end{figure*}

As noted above, the temporal flux variations seen here are often accompanied by a
change in the optical to X-ray spectrum. This can be seen clearly from the optical
to X-ray SED in Figure \ref{fig:fig4} where we show a collection of optical to X-ray 
SED, representative of spectral changes seen during this period. It also shows the 
best-fit spectral model for the X-ray along with the 1\(\sigma\) region. 

The most notable aspect of this strong spectral evolution is the drastic evolution 
shown by the high-energy-end of this broadband component responsible for the softer
X-ray spectrum. This behavior is clear from focusing on the SED of duration corresponding
to the shaded regions in Figure \ref{fig:mwlc}. Though there is a hardening of optical
to UV spectrum, and often an increase in the level of emission as well, the X-ray spectrum
remains similar except for a change in the level of emission e.g. SEDs of MJD 58054.3
and 58174.2 in the 1st panel of Figure \ref{fig:fig4} and SED of MJD 58994.3 and 59006.7
in the last panel. These trends and behaviors are similar to the one seen during its 
highest reported X-ray activity in 2016--2017 \citep{2018MNRAS.479.1672K}.

\begin{figure}%[ht]
\centering
 \includegraphics[scale=0.42]{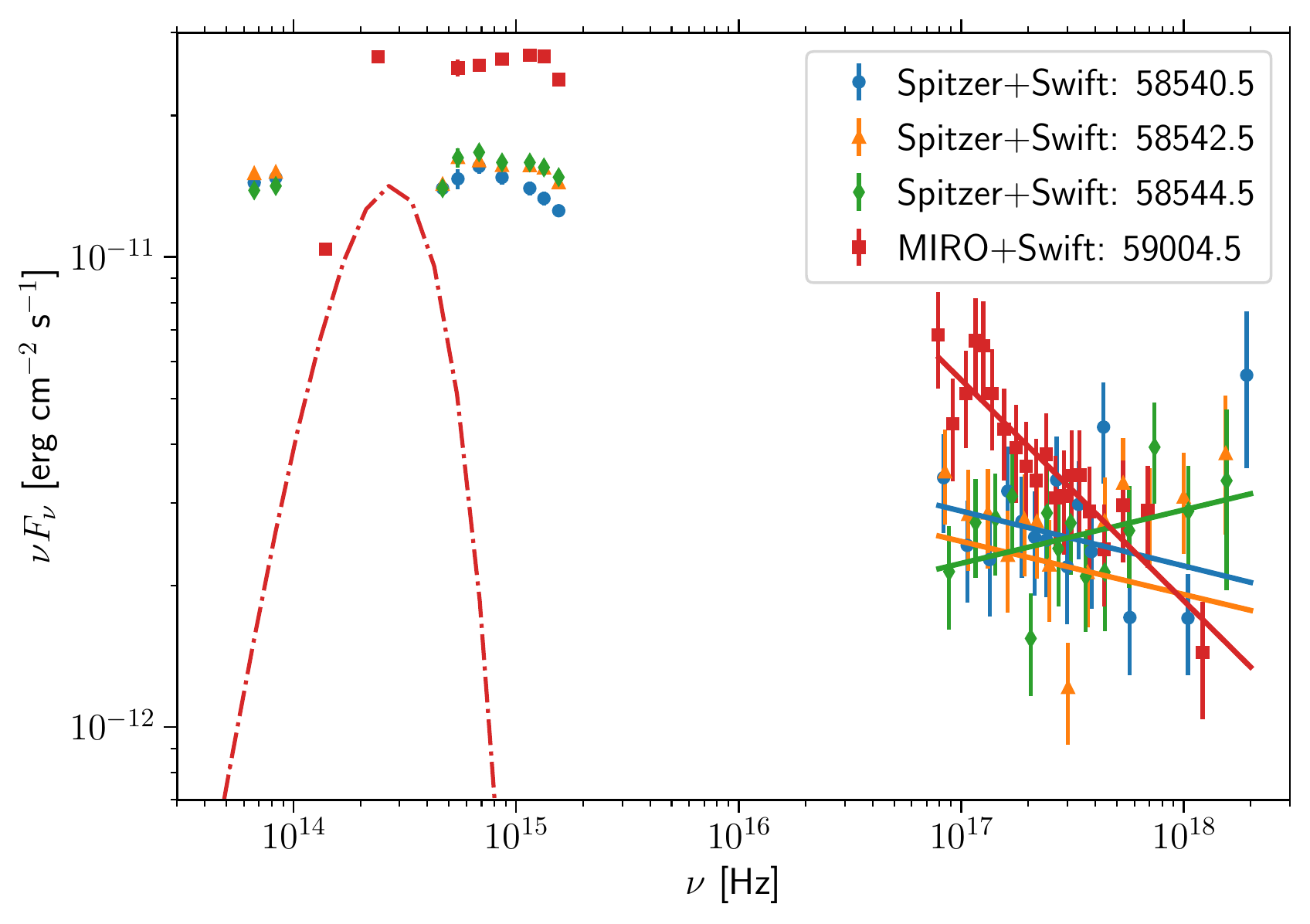}
\caption{IR to X-ray SED of OJ 287 when NIR (Spitzer/MIRO) and Swift have oberved
within that MJD. The MJDs mark the mid point and the solid lines are the best-fit
X-ray model. The red dashed curve is the standard accretion disk spectrum of a
$1.8\times 10^{10}~M_\odot$ SMBH with outer radius truncated at 9 gravitational
radius.}
\label{fig:irSED}
\end{figure}

Figure \ref{fig:irSED} shows the NIR to X-ray SED on June 4, 2020 (MJD 59004.5).
The NIR data indicates a strong spectral change from the well-known power-law
optical to NIR spectrum of OJ 287, similar to the change observed during 2015
activity of OJ 287 \citep{2018MNRAS.473.1145K,2020Galax...8...15K}. Since
the NIR-optical change reported during 2015 observation was consistent with 
a standard accretion-disk spectrum \citep{2018MNRAS.473.1145K}, here
too we again explored this aspect.  As per 
\citep[Table 3;][]{2018ApJ...866...11D},
the 2019 impact is at $\sim$9 gravitational radius of the primary SMBH of mass
$1.8\times10^{10}~M_\odot$, where $M_\odot$ is the mass of the Sun. Plotting 
the standard accretion disk spectrum of a  SMBH with the 
outer disk radius truncated at the impact radius suggest that this sharp NIR
spectrum again could be the disk emission (see also \S\ref{subsec:joint}). 

To track whether this change of NIR state was like this even before, we also used
the Spitzer IR observation presented in \citet{2020ApJ...894L...1L}. Only the last
measurement on September 6, 2019 (MJD 58732.7) indicates a hint of a possible change in
NIR spectral state \citep[Table 1;][]{2020ApJ...894L...1L}. However, the lack
of IR/NIR data thereafter prohibits any further conclusion whether the IR
spectral change was persistent, like the way it was
in 2015 outburst \citep{2018MNRAS.473.1145K} and is present since then or transient. For the overlapping days of Swift and Spitzer's
observations, the IR to X-ray SEDs is shown in Figure \ref{fig:irSED}.

\subsubsection{Joint Spectral modeling}\label{subsec:joint}

\begin{figure}
\centering
\includegraphics[scale=0.3, angle=0]{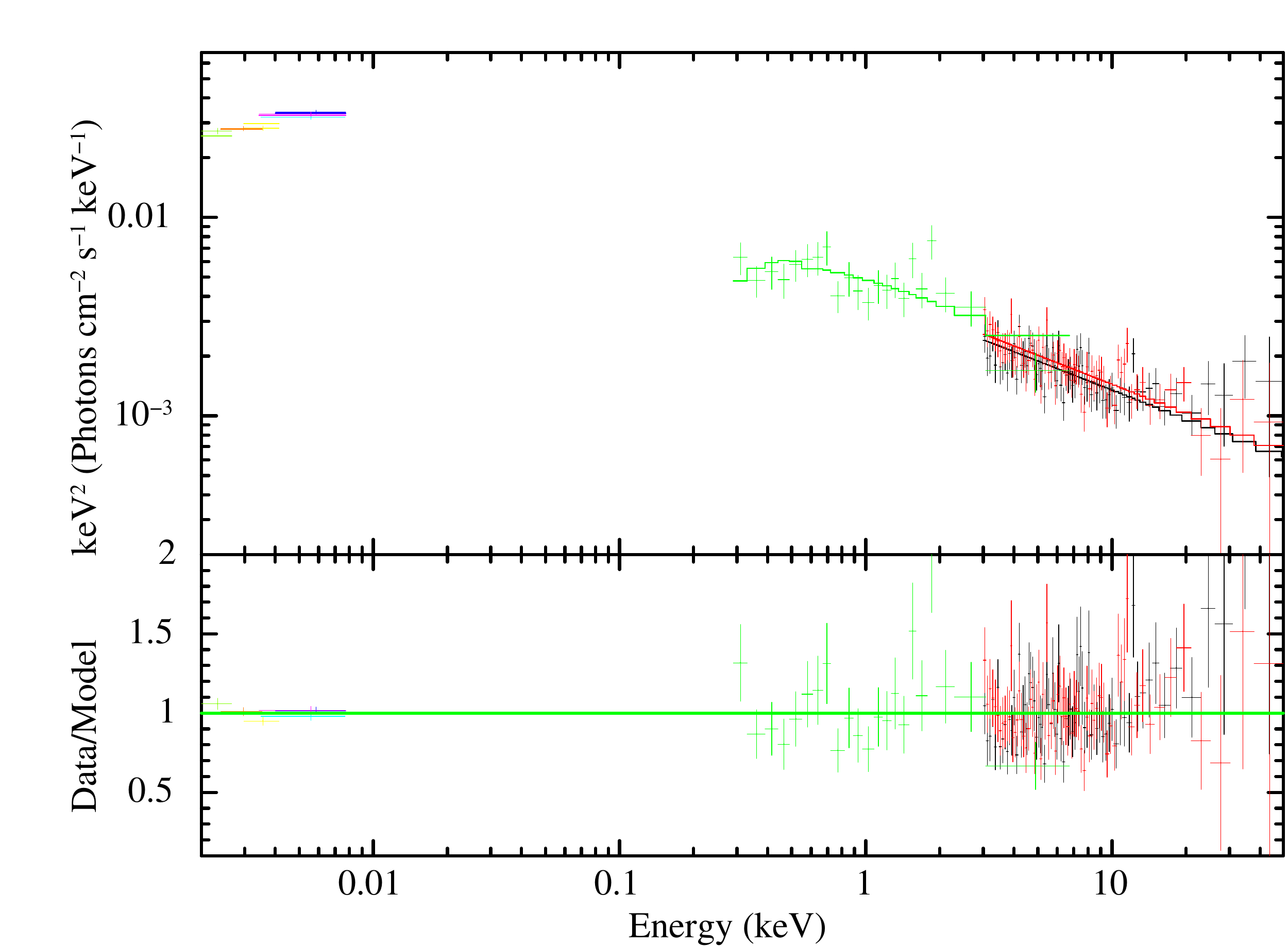}
\caption{Joint fit to the contemporaneous Swift (XRT: green and UVOT: six left most
data points) and NuSTAR data (FPMA: Black, FPMB: Red). The best-fit model, data and residuals are shown for the Model: {\tt zredden$\times$redden$\times$tbabs$\times$bknpower}
{(ref \ref{subsec:joint})}.}
\label{fig:lp_bpl}
\end{figure}

OJ 287 was also observed with {\it NuSTAR} on May 4, 2020 (MJD 58973) and
the joint {\it XMM-Newton}+{\it NuSTAR} X-ray spectral fit is presented
in \citet{10.1093mnrasl}. Here, we explored the joint optical to X-ray fit using
the {\it Swift} and {\it NuSTAR} data in XSPEC. We first fitted a PL model
to the \textit{Swift}-XRT and {\it NuSTAR} data in 3-50 keV range. We then extended the 
lower energy range to 0.3 keV and modified the PL model for Galactic absorption
via {\tt tbabs} and fixed the column density to $N_{H}=2.38 \times10^{20}$ $cm^{-2}$.
Additionally, we added a constant function {\tt constant} to cross-calibrate
various instruments and fixed it to unity for the FPMA module and free for others.
The fit resulted in $C_{stat}/dof$ $=886.1/994$, where {\it dof} is degree of freedom.
There were no significant features in
the residuals for the PL model. We found a PL photon index $\Gamma=2.39\pm0.04$
in the 3-50 keV. We then added XRT data down to 0.3 keV. Extending the PL fit below
3 keV showed some excess. So we first froze the above PL components and fitted
this feature by an additional {\tt PL} model. This fit resulted in a $C_{stat}/dof$ $=1059.9/1186$. The best-fit {\tt PL} index was found to be $\Gamma=3.28_{-0.47}^{+1.02}$.
Finally, we let the frozen parameters of the {\tt PL} model vary and the fit resulted
in $C_{stat}/dof$ $=1058.7/1184$. 

\begin{table}
\centering
\caption{Best fit parameter for joint Swift + NuSTAR data using Model: {\tt zredden$\times$redden$\times$tbabs$\times$bknpower}. `f' stands for fixed value.}% and Model-2:{\tt zredden$\times$redden$\times$tbabs$\times$(zbbody $+$bknpower)}. `f' stands for fixed value.}
 \begin{tabular}[c]{ccc}
\hline 
Model Parameter & Model component \\ \hline %& Model-2  \\ \hline
Intrinsic Reddening($E_{B-V}$) & $\le4.1\times10^{-4}$ \tabularnewline% & $\le3\times10^{-4}$  \tabularnewline
Reddening ($E_{B-V}$) & $0.0241(f)$  \tabularnewline%& $0.0241(f)$  \tabularnewline

\hline  
$\Gamma_{BPL}^{1}$ & $1.861\pm0.002$  \tabularnewline%&$2.26_{-0.14}^{+0.07}$  \tabularnewline
$E_{br}(keV)$ & $0.0067_{0-0.0007}^{+0.000003}$  \tabularnewline%& $0.049_{-0.044}^{+0.62}$  \tabularnewline

$\Gamma_{BPL}^{2}$ & $2.434\pm0.004$  \tabularnewline%& $2.40\pm0.03$  \tabularnewline
 
$N_{BPL}(10^{-2})$ & $6.56\pm0.06$  \tabularnewline%& $0.5_{-0.2}^{+0.8}$  \tabularnewline
Statistic $(C_{stat}/dof)$ & $1082.9/1189$  \tabularnewline%& $1074.3/1187$  \tabularnewline
\hline
\end{tabular}
 \label{tab:jointFit}
\end{table}

The above fit indicates two statistically different photon spectral indices ($
\Gamma_{3-50 keV} \sim 2.4$ and $\Gamma_{0.3-3 keV}\sim 4$), suggesting that    a broken
PL model could be fitted in the entire 0.3-50 keV band. We replaced the two PL
models by a broken PL model ({\tt bknpl}). The fit resulted in $C_{stat}/dof$ $=1059.5/1184$.
We also used the LP model: {\tt tbabs$\times$logpar} and the fit resulted in
$C_{stat}/dof$ $=1057.7/1185$.

We then added the UV-optical data corresponding to the X-ray observation. Since 
optical-UV and X-ray have different spectral indices, we used a {\tt bknpl} model
for the whole optical to X-ray data. Further, for reddening correction,
we added the {\tt redden} model and fixed the reddening factor due to our Galaxy 
with $E_{B-V}=0.0241$ \citep{2011ApJ...737..103S}. We also added a redshifted reddening
model {\tt zredden} to modify reddening intrinsic to the OJ 287. When we extrapolated
the best-fit model fitted on the X-ray data down to optical data, we got large residuals.
Refitting the entire optical to X-ray spectrum resulted in $C_{stat}/dof$ $=1099.9/1189$.
An expected black body was added to check any improvement in the fit. With two free
parameters, the redshifted black-body {\tt zbbody} model did not improve
the fit statistic. The best-fit model, data, and residuals are shown
in Figure~\ref{fig:lp_bpl} and the corresponding parameters are listed in
Table~\ref{tab:jointFit}.

\section{Discussion}\label{sec:discussion}
The MW spectral and temporal study presented here covers a duration of 3 years,
from June 16, 2017 (MJD 57920) to June 20, 2020 (MJD 59020). This corresponds to
the duration after the end phase of the highest reported X-ray flux state of OJ 287 in
2017 \citep{2017ATel10043....1G,2018MNRAS.479.1672K} to the end of the second-highest X-ray flux state,
observed in April 2020 \citep{10.1093mnrasl}. The coverage thus provides inhibited
access to the evolution and track of the emission state of the source after the
highest X-ray flux state which was due to a new broadband emission component
\citep{2018MNRAS.479.1672K}, thereby, allowing a view of this elusive and
rare behavior of the source \citep[see also][]{10.1093mnrasl}.

\subsection{Flux evolution}
The period considered here shows intense optical to X-ray activity, with most of the
variations associated with a change in the optical-UV and X-ray spectral state of
the source (ref Fig. \ref{fig:mwlc} bottom plot). The optical to X-ray variations
are correlated (in general) and simultaneous within the observational cadence except
for two durations (shaded regions in Fig. \ref{fig:mwlc}) when the flux changes show
a complex trend. Though the variations are anti-correlated during both the shaded
periods, in the blue-shaded period (MJD: \(\sim\) 58040 -- 58085), X-ray shows a
flare while optical-UV flux density is towards the minimum of the current duration.
Contrary to this, in the red shaded region (MJD: \(\sim\) 58980 -- 59005), the
optical-UV is at maximum while X-ray is quite close to the minimum of the considered
duration. Further, the optical-UV flux-flux variations are linear (right panel Fig.
\ref{fig:fig2}) but not the optical/UV with X-ray except for a few. 

In the \(\gamma\)-ray band at {\it Fermi}-LAT  
energies, the analysis around the high X-ray flux state indicates that the source
is rather weak even over weekly intervals with only a few detections if we use an LP
spectral profile -- the default 4FGL model of the source (ref. Fig. \ref{fig:lat}
top panel). Changing the source model to PL result in no detection at all over
weekly time-bins. The
no detection at all in the PL model could be a combined effect of relatively low
flux state of the source \citep[e.g.][state 3 SED]{2013MNRAS.433.2380K} and the non-PL MeV-GeV spectrum (bottom panel Fig. \ref{fig:lat}). Surprisingly,
LAT detections suggest flare much before the observed second brightest
X-ray flare.

The optical to X-ray flux variation trends are quite similar to the one observed
during the 2016--2017 activity except for the amplitude and duration of the activity
-- peak X-ray flux being smaller \citep{10.1093mnrasl} and also a quite shorter duration
of activity during the current period \citep{2018MNRAS.479.1672K,10.1093mnrasl}.
Like the 2016--2017 activity, some period of current duration also suggests lag
between UV and X-rays \citep{galaxies8030058}.

\subsection{Spectral evolution}
The spectral evolution trends are much more complex. Broadly, we can see an
anti-correlation between the optical-UV and X-ray emission spectra (ref. Fig
\ref{fig:mwlc} bottom plot) but this is clear only for the latter half period (MJD
$\gtrsim$ 58500). For optical-UV, the flux is anti-correlated with the optical-UV
spectrum throughout i.e. flux increase is associated with hardening of the
optical-UV  spectrum and vice-versa. For X-ray, on the other hand, 
the trend is clear only in the latter half (after MJD \(\sim\) 58500) and is opposite
of optical-UV i.e. an increase in X-ray flux is associated with a softening of the
spectrum and vice-versa. This anti-correlation between optical-UV and X-ray
spectrum has been reported before \citep{2017MNRAS.468..426S}, however, the current
spectral state is very different. Looking at the optical to X-ray SED (ref. Fig. \ref{fig:fig4})
evolution, one can see that the high-energy-end of the component responsible for
optical-UV hardening is very dynamic and often becomes too steep -- like
a cutoff (ref. panels 1, 5, and 7 of Fig. \ref{fig:fig4}), making a negligible
contribution at X-ray energies. Hence, we see a softening and flux increment in the
latter half when this high-energy-end is not that steep.

At LAT \(\gamma\)-ray energies as well, an appreciable hardening in the MeV-GeV
spectra is visible (\(\alpha \sim 2\), Fig. \ref{fig:lat}) around the highest X-ray
flux state. This is also visible in the MeV-GeV SED, showing a hardening over \(\sim
1 \) GeV, contrary to its characteristic simple PL SED \citep[e.g.][]
{2010ApJ...716...30A,2013MNRAS.433.2380K}. Spitzer and Swift SEDs show that the
spectral shape of IR SED is consistent with the generic SED of OJ 287 where the 
low energy peak is around the NIR band. However, MIRO observation shows a departure
from its typical smooth PL NIR to optical SED \citep[e.g.][]{2020Galax...8...15K}, 
similar to the one observed during the 2015 outburst \citep{2018MNRAS.473.1145K}.
 Within the disk-impact binary SMBH model \citep{2018ApJ...866...11D}, this
NIR spectrum can be reproduced fairly well by the standard accretion disk
spectrum of the primary SMBH with outer disk truncated at the impact point 
i.e. 9 gravitational radius (Fig. \ref{fig:irSED}). A thermal component
(T \(\sim 3\times10^4\) K) is also hinted by the joint fitting of optical
to X-ray data from Swift and NuSTAR with a broken power-law model (Fig. \ref{fig:lp_bpl} and Table \ref{tab:jointFit}). This temperature
is similar to the one 
found during the 2015 outburst, responsible for the break in the NIR and optical
SED \citep{2018MNRAS.473.1145K}. Alternately, a thermal bremsstrahlung can 
also reproduce this spectral change \citep{2020MNRAS.498.5424R} as claimed in the
disk-impact model of OJ 287 central engine \citep{2018ApJ...866...11D}.

The observed spectral changes reported here, from optical to X-ray as well as
MeV-GeV energies, unambiguously establish the changes to a new HBL like broadband
emission component, similar to the one observed during the highest reported X-ray
flux and activity state of OJ 287 in 2016--2017 \citep{2018MNRAS.479.1672K}. 
The spectral evolution presented here shows that the HBL like emission
component seen during
2016 -- 2017 activity has not vanished completely as the X-ray spectrum has not 
reverted to the known power-law spectrum with an index of \(\Gamma \sim 1.5 - 1.6\)
as seen during the shaded blue period (MJD \(\sim\) 58040 -- 58085, Fig. \ref{fig:mwlc})
when optical-UV is minimum. Following the trend presented in \citet{2018MNRAS.479.1672K}
with results here show that it
had continued to weaken till it has reached the hard X-ray spectrum around MJD \(58050\) when the UVOT measurement touched the lowest in the current episode.
It remains so for about \(\sim 50\) days and then this X-ray spectrum starts
to become softer while optical-UV becomes harder, marking the revival of this
component. After this, it remains present more or less, eventually becoming
stronger, reaching the 2nd highest reported X-ray flux in end-April 2020
\citep[MJD \(\sim 59965\);][and references therein]{10.1093mnrasl}.

\subsection{Traditional HBL vs OJ 287 new Broadband Emission Component}
Though the overall optical to MeV-GeV spectrum points to an HBL like broadband
component with the synchrotron peak in UV region, the high-energy-end
(after UV bands) of this component in the case of OJ 287 appears dynamically
very different and uncharacteristic of traditionally known HBL blazars. The X-ray
photon spectral index (\(\Gamma/\alpha \gtrsim 2.5 - 3\); 0.3--10 keV) around the
peak flux is typical of well-known HBLs that have synchrotron peak around UV energies
\citep[e.g. Mrk 421; ][]{2016ApJ...819..156B}, but it is drastically
variable in OJ 287 and often becomes too steep -- more like a spectral cutoff
(ref Fig. \ref{fig:fig4}; spectra of shaded periods of Fig. \ref{fig:mwlc}). 

 Apart from the similarity of observed spectral variations during both the activity
periods (2016--2017 and current) driven by an extremely soft X-ray spectrum, another
correlation seems to be the strength of the modified MeV-GeV spectrum and the peak X-ray
flux exhibited by the source. The modified MeV-GeV spectrum during the current high
activity period is similar to the one observed during 2016--2017 activity but the
level of emission is lower (ref Fig. \ref{fig:lat}, bottom panel), as is the peak
X-ray flux which
is lower in the current episode compared to its peak during 2016--2017 activity.
Further, based on the MeV-GeV spectrum, VHE emission is expected but there is
no such report for the current activity. Given the spectral similarity of optical
to X-ray and MeV-GeV emission, some inferences can be drawn regarding the lack
of VHE based on the 2016--2017 activity \citep{2017ICRC...35..650B,
2018MNRAS.479.1672K}. In general, the observed VHE spectrum is expected to be steeper
due to attenuation caused by the extra-galactic background light \citep[EBL;][]
{2011MNRAS.410.2556D}.
The VHE spectrum during the 2016--2017 activity is consistent with a PL photon
spectral index ($\rm \Gamma_{VHE}$) of $3.58\pm0.32$ \citep{2017ICRC...35..650B}. 
Correcting this for EBL leads to an intrinsic $\rm \Gamma_{VHE} \sim 2.58\pm0.32$, 
similar to the X-ray photon spectral index ($\Gamma_X \sim 2.6$) during and around
the peak X-ray flux \citep{2017ATel10043....1G,2017ICRC...35..650B,2018MNRAS.479.1672K}.
Since $\Gamma_X$
during and around the peak X-ray flux during the current activity is $\sim 2.8-3$,
assuming a similar intrinsic VHE spectrum would lead to a $\rm \Gamma_{VHE} \sim 3.8-4$
due to EBL attenuation. This steepness and a low MeV-GeV flux in the current 
episode combined with the empirical fact that the amplitude of high-energy
hump in HBL is lower than the low-energy hump could be one of the reasons for non-detection of the source at VHE energies.
Though there can be other reasons, it should be noted that even during the 2016--2017
activity, the VHE was detected quite late \citep{2017ICRC...35..650B} -- almost by
3 -- 4 months with respect to the commencement of such optical to X-ray spectral state \citep{2017IAUS..324..168K,2018MNRAS.479.1672K}.

Another supporting evidence is that during the 2016--2017 high
X-ray activity, the X-ray
spectrum was quite steeper (\(\Gamma \sim 3\)) and fluctuating in the beginning and
the end while it was relatively flatter (\(\Gamma \sim 2.6\)) and stable during
the duration of VHE detection \citep[MJD $\sim$ 57750 -- 57840;][]{2017ICRC...35..650B}.
A similar pattern can be noticed during the high X-ray flux of the current activity
episode. We would also like to point out that in the broadband SED modeling in 
\citet{2018MNRAS.479.1672K}, the VHE spectrum is not corrected for EBL absorption
\citep{2011MNRAS.410.2556D} and thus, though the conclusion and inferences remain valid, the model parameters
for the 2nd zone are not. In fact, as shown above, the EBL corrected
VHE spectrum reported by the VERITAS \citep{2017ICRC...35..650B} is consistent
with the X-ray spectral index, indicating that inverse Compton scattering responsible
for VHE emission happened in the Thomson scattering regime. 

A noticeable and important difference between the current MeV-GeV spectrum from
LAT with the one seen during the 2016--2017 activity is that the traditional
MeV-GeV spectrum is comparatively stronger and clearly visible in the LAT spectrum
of the current duration (Fig. \ref{fig:lat}).

\subsection{Unique Trends of the New Spectral State}
There are a few unique spectral and temporal features common during both the episodes
of high X-ray activity driven by the new HBL like emission component (shaded regions
in Fig. \ref{fig:mwlc}). The most prominent being a minimum in optical-UV when the
X-ray is relatively high or probably at a base level flux with a hardened X-ray
spectral state, followed subsequently by a low in X-ray but a high one in optical-UV
(the red shaded region, Fig. \ref{fig:mwlc}), after the peak X-ray flux
\citep[see also \citet{10.1093mnrasl}]{2018MNRAS.479.1672K}. The time separation between the two, however, is different. It was $\sim$ 10 days in 2016--2017
activity \citep[Table 1 and associated texts;][]{2018MNRAS.479.1672K} while it is $\sim$ 30 days during
the current activity episode. Another feature could be a minimum in optical-UV but
a base or flaring in X-ray (the blue shaded region, Fig. \ref{fig:mwlc}). Unfortunately,
unless we catch this feature in future observations, it will remain ambiguous given
our lack of predicting its time of occurrence and our inability to continuously
monitor an astrophysical source. 

The other unique feature is the drastic evolution of the high-energy-end of this
new emission component which goes from a PL shape to a very steep, more like
a cutoff
within a few days. This steepness is the reason for dramatic flux dimming in X-ray 
without much change in optical-UV emission level and spectral shape (Fig.
\ref{fig:fig4}, e.g. SEDs labeled 58965.7 and 58994.3).

\subsection{The New Spectral State and $\sim$12-yr optical QPOs}

Though X-ray observations of OJ 287 before the year 2000 are only a few, mainly around
the claimed $\sim$ 12-year outbursts and hence any long-term inferences derived using
them may be biased, these records do indicate such soft X-ray spectral state within
a few years around the peculiar
$\sim$ 12-year outburst \citep[and references therein]{2001PASJ...53...79I,
2020Galax...8...15K}. However, the 2005 and 2007 outbursts do not show such X-ray
state or even hardening in optical-UV. For the latest two of these outbursts, happened
in 2015 and 2019, such extremely soft X-ray spectral state appeared respectively
\(\sim 10\) and \(\sim 9\) months (this work; see also \citet{10.1093mnrasl}) after the claimed disk-impact outbursts \citep{2018ApJ...866...11D,2020ApJ...894L...1L} within the limits of observational
records in the literature. The records and timing, thus, clearly indicate a
connection between the two and also point to the peculiarity of these outbursts
compared to typical OJ 287 outbursts \citep[see also][]{2020Galax...8...15K}.

In addition to predicting and explaining the recurrent $\sim$ 12-year optical
outbursts, the disk-impact binary SMBH model also predicts activity as a result
of perturbation and tidal effects \citep{2009ApJ...698..781V,2013ApJ...764....5P}.
The current observational results
-- strong and rapid variability as well as the broadband spectral features, on the 
other hand, unambiguously indicate the emission to the jet as also pointed
out by \citet{10.1093mnrasl} based on variability, spectral evolution, and 
high polarization \citep{2020ATel13637....1Z}. Thus, whether the
observed variations are an outcome of these effects reaching the jet, thereby
indicating a jet-disk connection -- an unsolved and highly debated research topic,
or something else
is still unclear at the moment.

\section{Summary}
We presented a detailed study of the spectro-temporal evolution and broadband 
spectral behavior of OJ 287 between May 27,
2017 to June 10, 2020. The period corresponds to the end phase of the highest reported
X-ray activity of OJ 287 in 2016--2017 to the end phase of its second highest X-ray
flux state that occurred in end-April 2020. This continuation provides the view of
evolutionary fate of the new HBL like broadband emission component that was responsible
for the 2016-2017 extreme optical to X-ray variability. A summary of observed
spectral and temporal behavior and inferences is as follow:

\begin{itemize}
 \item Except for two period of duration $\sim$ 50 and $\sim$ 10 days, optical to
 X-ray light curves exhibit strong, correlated variability
 for the rest, associated with spectral changes, especially at X-ray energies.
 At LAT \(\gamma\)-ray energies, the light curve around highest X-ray flux reveal
 that the source is weaker, with only a few detection. For optical-UV emission,
 the spectrum and flux variations are correlated while the trend is opposite for
 variations in X-ray band.
 
  \item IR to MeV-GeV spectra show very different spectral state compared to its
 well-known spectrum, indicating new spectral features in all the bands. The
 resultant broadband SED (IR to MeV-GeV) establishes the observed spectral and temporal
 variations to a highly dynamic HBL like emission component, similar to the one
 seen during the 2016--2017 high X-ray activity period of OJ 287. 
 
 \item The optical-UV and X-ray flux variations are anti-correlated for the 
  above mentioned two periods (exhibiting odd behaviors), and are mainly
  due to the steepening of the high-energy end of the HBL emission component.
  
 \item The new emission component responsible for the strong optical
 to X-ray activity during the current and the 2016--2017 period though appears
 similar to the well-known blazars HBL component, its high-energy end is
 extremely dynamic and become too steep within a day and thus, untypical in
 this sense.
 
 \item The observation of a dip in X-ray but a high optical-UV flux after 
 the highest X-ray flux state during the current and 2016--2017
 activity indicates this behavior to be a characteristic feature of this new
 broadband emission component.
\end{itemize}

\section*{Acknowledgments}
We thank the anonymous referee for his/her invaluable inputs. PK acknowledges support from ARIES A-PDF grant (AO/A-PDF/770). MP thanks the financial
support of UGC, India program through DSKPDF fellowship (grant No. BSR/ 2017-2018/PH/0111).
N Kalita acknowledges funding from the Chinese Academy of Sciences President’s
International Fellowship Initiative Grant No. 2020PM0029. ACG is partially supported
by Chinese Academy of Sciences (CAS) President’s International Fellowship Initiative
(PIFI) (grant no. 2016VMB073). EMGDP acknowledges support from the Brazilian funding
agencies FAPESP (grant 2013/10559-5) and CNPq (grant 308643/2017-8 ). MFG acknowledges
support from the National Science Foundation of China (11873073). The work at Physical
Research Laboratory is supported by the Department of Space, Govt. of India. This
research has made use of the XRT Data Analysis Software (XRTDAS) developed under the 
responsibility of the ASI Science Data Center (ASDC), Italy.

\vspace{5mm}
\facilities{Swift (XRT and UVOT), NuSTAR, Fermi-LAT, MIRO}

\software{HEASOFT (\url{https://heasarc.gsfc.nasa.gov/docs/software/heasoft/}),
IRAF \citep{1986SPIE..627..733T,1993ASPC...52..173T}, 
Matplotlib \citep{Hunter2007}, Gammapy \citep{2017ICRC...35..766D,2019A&A...625A..10N},  
Gnuplot (version: 5.0; \url{http://www.gnuplot.info/})}

\bibliographystyle{aasjournal}
\bibliography{oj287_2020F_rv2}

\end{document}